\documentclass[11pt]{article} 

\usepackage[utf8]{inputenc} 
\usepackage[T1]{fontenc}

\usepackage{geometry} 
\usepackage{graphicx} 

\usepackage{booktabs} 
\usepackage{multirow} 
\usepackage{makecell} 
\usepackage{array} 
\usepackage{paralist} 
\usepackage{verbatim} 
\usepackage{subfig} 
\usepackage{authblk} 
\usepackage{amsmath}  
\usepackage{amssymb}

\usepackage{fancyhdr} 
\usepackage{sectsty}
\allsectionsfont{\sffamily\mdseries\upshape} 

\usepackage[nottoc,notlof,notlot]{tocbibind} 
\usepackage[titles,subfigure]{tocloft} 

\usepackage{multirow} 
\usepackage{makecell} 

\title{Article title}
\title{Optical free space extreme learning machine for the implementation of emergent complex systems}

\author[1]{E. Moreno}
\author[1,*]{F. Soldevila}
\author[1,*]{D. Torrent}
\affil[1]{GROC, UJI, Institut de Noves Tecnologies de la Imatge (INIT), Universitat Jaume I, 12071 Castelló, Spain}
\affil[*]{Corresponding authors: fsoldevi@uji.es, dtorrent@uji.es}
\date{} 

\begin{document}
\maketitle
\abstract{Cellular automata conform a set of computational models which evolve with a reduced set of simple rules, yet still are able to show extremely complex emergent phenomena such as fractals and universal computation. Despite their apparent simplicity, they have shown great potential in simulating natural systems and solving challenging computational tasks such as classification and image generation. Instead of implementing cellular automata purely at the software level, it is desirable to design novel analog computing platforms that physically evolve following the automata's underlying rules, thereby reducing power requirements and latency. Here, we introduce an optical extreme learning machine for the simulation of a wide range of cellular automata. Our system operates in free space, and uses a spatial light modulator to encode the evolution rules of the system, while coherent wave propagation performs the corresponding computations. Our results demonstrate a simple, fully-programmable, cost and power efficient, and easy to build and align platform for the implementation of a wide range of complex computational systems such as elementary cellular automata, Conway's Game of Life, and two-dimensional Turing machines.}

\section*{Introduction}
Built on top of transistor technology and the Von Neumann architecture, modern digital computing systems excel at solving algorithmic problems with explicit rules and sequential instructions. However, when faced with the task of simulating natural systems, typically formed by complex assemblies of interconnected elements that collectively evolve following a given set of rules which might not even be known, this paradigm encounters energy consumption and information bandwidth bottlenecks \cite{gholamiAIMemoryWall2024}. Interestingly, some of the most relevant open research questions today, like understanding the human brain, energy generation, or climate and environmental modeling, belong to this second group of problems. While there have been huge advances in these fields by developing novel digital computing architectures that mimic the structure of the systems under study (for example, the development of modern artificial neural networks), the physical substrate used to implement them ultimately limits their capabilities \cite{gholamiAIMemoryWall2024, leisersonTheresPlentyRoom2020}. A different paradigm to tackle these bottlenecks is to develop novel hardware that, instead of trying to digitally implement the underlying natural rules of a complex system using transistors and a set of algorithms, directly maps the system’s behavior using the physical evolution of its components. 

Photonic-based systems have shown very promising capabilities in the design and fabrication of hardware under this new computing paradigm. First, a large number of photons can be generated and transmitted using low energy levels in small volumes, enabling the miniaturization of devices. Second, they are able to multiplex large amounts of information using polarization, wavelength, phase, and/or orbital angular momentum, which allows to create high bandwidth systems. Moreover, optical state transitions typically happen at extremely short timescales, highly reducing latency bottlenecks. Taking advantage of these properties, in the fields of image processing and machine vision, optical artificial neural networks (ANN) have demonstrated state-of-the-art performances on recognition \cite{liangHighclockrateFreespaceOptical2026, geAllopticalLogicProcessing2026}, classification \cite{svedInversedesignedNanophotonicNeural2026, liModelfreeOpticalProcessors2026}, and image generation \cite{oguzOpticalDiffusionModels2024, chenOpticalGenerativeModels2025, liModelfreeOpticalProcessors2026} tasks. Likewise, it has been shown that complex optimization problems can be efficiently solved using optical Ising machines \cite{pierangeliLargeScalePhotonicIsing2019, wuMonolithicallyIntegratedOptical2025}, and the use of optical reservoirs can be used to forecast the evolution of chaotic systems \cite{rafayelyanLargeScaleOpticalReservoir2020, wangOpticalNextGeneration2025}.

Since its introduction in the 1940s, cellular automata (CA) have been widely used as a platform to study how complex behavior can emerge on systems governed by small sets of simple rules \cite{neumannTheorySelfreproducingAutomata1966}. CA consist of an ensemble of cells arranged in a regular lattice, and the state of each individual cell is calculated based on a set of rules and the states of the cell and its neighbors. While apparently simple, research has shown that CA are able to replicate extremely complex behavior such as fractals, chaos, and even universal computation \cite{wolframUniversalityComplexityCellular1984}. Recently, the use of generative tools has introduced the design of neural CA, where the rules that define the evolution of the system are designed so the behavior of the CA directly matches the task at hand. This allows to model natural systems and tackle real-world problems such as traffic flow \cite{liuCellularAutomataTraffic2019}, wildfire spread \cite{zhengForestFireSpread2017}, image generation \cite{mordvintsevGrowingNeuralCellular2020, pajouheshgarNeuralCellularAutomata2026}, classification \cite{liDeepLearningPhotonic2024}, and video game design \cite{earleIlluminatingDiverseNeural2022}. One striking characteristic of some CA of interest is that they are computationally irreducible (i.e., the only way to evaluate their state at a given time is to calculate all the preceding states), which makes it desirable to find novel hardware designs to evaluate them without the high-power demands of their implementation on conventional computers.

Here, we introduce a free space optical extreme learning machine (ELM) for the calculation of different classes of CA. The biggest benefit of an optical ELM is its experimental simplicity, as its fundamental building blocks are currently present in almost every optics lab: a coherent laser source, a spatial light modulator (SLM), a few lenses, and a digital camera. Compared to other physical implementations of CA \cite{ liDeepLearningPhotonic2024, rothemundAlgorithmicSelfAssemblyDNA2004, itohMemristorCellularAutomata2009, liPhotonicElementaryCellular2023, zhangLargescaleOpticalProgrammable2024, zhangPhotonicLogicTensor2025}, our approach is fully programmable, easy to build, and inexpensive. At the core of our system lies a liquid crystal on silicon SLM, which we use to spatially modify the field of a coherent beam of light and encode information. The modified field distribution is then Fourier transformed via the use of a single lens, and the resulting intensity at the detector plane is measured with a pixelated detector. We show how, using this codification scheme, every detector pixel can act as a computational node in an ELM, enabling to reproduce a broad range of complex behaviors by implementing Elementary Cellular Automata (ECA), Conway’s Game of Life (CGoL), and even two dimensional Turing machines. 

\section*{Results}
\subsection*{Principle}

\begin{figure}[h!]
	\centering
  	\includegraphics[width = 0.9\linewidth]{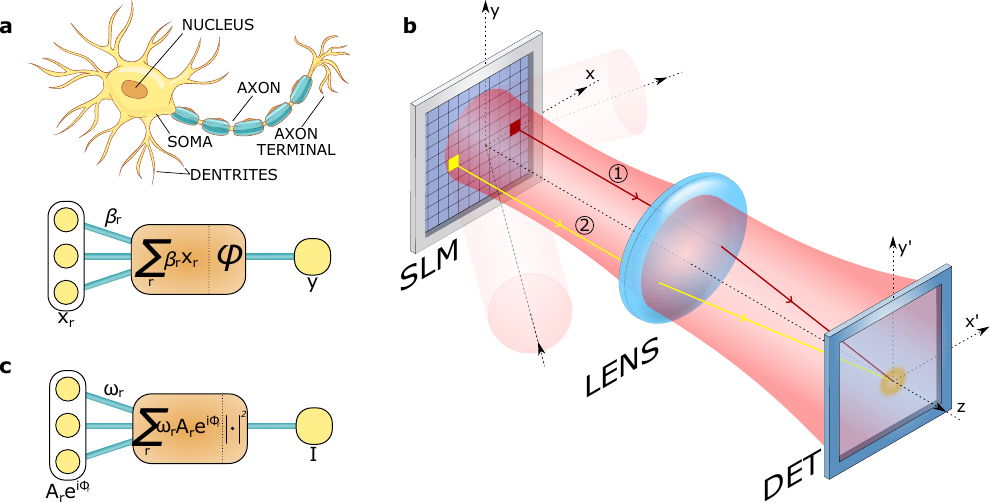}
  	\caption{Optical computing using free space optics. {\bf a} Comparison between a biological (top) and artificial (bottom) neuron. The soma of the cell in a neuron is able to launch an activation potential through the axon, which reaches the dentrites of neighboring cells, and thus the signal is transmitted through the network. ANs are designed following the same model: a series of inputs ($x_r$) is linearly combined via some weights ($\beta_r$), and after the application of a nonlinear activation function ($\varphi$), the output ($y$) is transmitted to other neurons. {\bf b} Conceptual view of an optical ELM. A collimated light beam illuminates a reflective SLM, which allows to control the complex amplitude and phase of the beam. Both the SLM and the detector (DET) planes are linked via a FT performed by a lens. Each pixel of the SLM can be thought as a small source with a controllable relative phase with respect to all the other pixels. After propagation, the intensity at a given pixel of the detector will be the result of the interference between all the individual waves generated by the SLM-lens system (for example, the control of the relative phases between the two highlighted pixels will change the intensity on the detector generated by the interference of rays \#1 and \#2). The phase control provided by the SLM allows to encode different operations in the resulting interference patterns measured on the detector. {\bf c} Computation node of the system. Contrary to conventional ANs, the computation node introduced here codifies the inputs on the complex field of a light beam, and the weights for the linear combination between fields are fixed and given by the geometry of the system. The non-linearity of the system comes both from the way the information is encoded (complex amplitude and phase of the field) and the measurement process on the detector (square modulus of the optical field).}
  	\label{fig:fig1}
\end{figure}

Before introducing the implementation of the optical ELM, we start with a brief description of an artificial neuron (AN), which constitutes the elementary building block of ANN architectures, and compare it with the computation node that forms the building block of our system. Inspired by biological neural circuitry, ANs consist of a series of inputs, which are weighted and linearly combined before being passed to an activation function. This was inspired by the dentrites, synaptic weights and activation potentials of biological neurons \cite{zouOverviewArtificialNeural2009}. Mathematically, they can be easily described by the following expression:

\begin{equation}
y_k = \varphi \left( \sum_{r=0}^{N-1}{\beta_{kr}x_{r} + b} \right),
\label{eq:artificial_neuron}
\end{equation}

where $\beta_{kr}$ are the weights that relate the $r^{th}$ input ($x_r$) to the $k^{th}$ output ($y_k$), $b$ is a bias term, and $\varphi$ is the activation function. Given a set of $N$ inputs (akin to the information introduced in the neuron via the dentrites), the output of the AN can either propagate through the neural architecture, or exit the system as an element of an output vector (see Fig. \ref{fig:fig1}a). When tackling complex problems such as image classification or chaotic series prediction, it is possible to design multilayer systems consisting of thousands or millions of ANs. Then, the value of each individual weight can be learned by using a training and test dataset and the backpropagation algorithm \cite{rumelhartBackpropagationBasicTheory1995}.

Many different physical systems can be used to replicate an AN and create ANN systems \cite{fernandoPatternRecognitionBucket2003, wrightDeepPhysicalNeural2022, chenPerformingOpticalLogic2024, huDiffractiveOpticalComputing2024, xiaNonlinearOpticalEncoding2024, zhangPhotonicLogicTensor2025, chenOpticalGenerativeModels2025}. Here, propose to implement different CA using an optical ELM \cite{pierangeliPhotonicExtremeLearning2021} based on free space optics which is simple, easy to build and train, and fully programmable. A conceptual view of the system is shown in Fig. \ref{fig:fig1}b. The fundamental idea is to use an SLM to code information in the complex field, optically perform the Fourier Transform (FT) of the field just by using a single lens, and measure the intensity of the field using a pixelated detector. At any position on the SLM plane, the complex field can be described as:

\begin{equation}
E(\vec{r}) = A(\vec{r}) e^{i\phi(\vec{r})},
\label{eq:field_slm}
\end{equation}

with $A(\vec{r})$ and $\phi(\vec{r})$ being the field amplitude and phase at the $\vec{r} = x\hat{x} + y\hat{y}$ spatial position. If we relate the SLM and detector planes via a FT using a single lens, the field at the detector plane can be expressed as:

\begin{equation}
E^{DET}(\vec{k}) 	\propto \iint_{\vec{r}} A(\vec{r}) e^{i\phi(\vec{r})} e^{-i\frac{2\pi}{\lambda f}(xx' + yy')} \,d\vec{r} ,
\label{eq:field_detector}
\end{equation}

where $E^{DET}(\vec{k}) $ is the field at the position $\vec{k} = x\prime\hat{x\prime} + y\prime\hat{y\prime}$ on the detector plane, and $\lambda$ and $f$ correspond to the wavelength and focal length of the lens, respectively. It is also possible to write a more compact form of the previous expression as:

\begin{equation}
E^{DET}_{k}  	\propto \int_{r} \omega_{k r} A_{r}  e^{i\phi_{r}} \,dr ,
\label{eq:field_detector_close_form}
\end{equation}

with $\omega_{k r} = e^{-i\frac{2\pi}{\lambda f}(xx' + yy')}$ acting as the kernel of the FT, and we introduce $r$ and $k$ as indices to indicate spatial positions in the SLM and detector planes, respectively. Last, given that the measured quantity is the intensity of the field, the result at any position of the detector plane is given by:

\begin{equation}
I_k = \left|\left| \int_r{w_{k r}A_r e^{i\phi_r}} \,dr \right|\right|^2,
\label{eq:computation_node_continuous}
\end{equation}

or, given that, in practice, both the spatial field modulation produced by the SLM and detector locations (pixels) are discrete:

\begin{equation}
I_k = \left|\left| \sum_{r}{w_{k r}A_re^{i\phi_r}} \right|\right|^2.
\label{eq:computation_node_discrete}
\end{equation}

It is worth noting both the similarities and differences between ANs (Eq. \ref{eq:artificial_neuron}) and the proposed computation nodes (Eq. \ref{eq:computation_node_discrete}). In both cases, the output depends non-linearly on all its inputs. However, while both paradigms share the fact that the inputs and outputs are related through a combination with some weights, the nature of these weights is radically different. In an AN, the weights ($\beta_{kr}$) are parameters that can be trained in order to make the system perform a given calculation (i.e. they can be considered as parameters of the system). In our approach, the weights ($\omega_{k r}$) are fixed and given by the geometry of the optical system and the FT properties of lenses (see Fig. \ref{fig:fig1}b-c). The existence of a bias term, while not initially considered in our system for simplicity reasons, could be added by implementing an extra light path arriving to the detector. While activation functions can be chosen at will on ANs, the computational nodes described here always operate with a square modulus nonlinear activation given by the intensity detection. However, this loss of degrees of freedom is compensated by the fact that the inputs can be encoded in the spatial distribution of amplitudes and phases with the SLM, which makes possible to implement different ANN architectures, such as ELMs.

\begin{figure}[h!]
	\centering
  	\includegraphics[width = 0.9\linewidth]{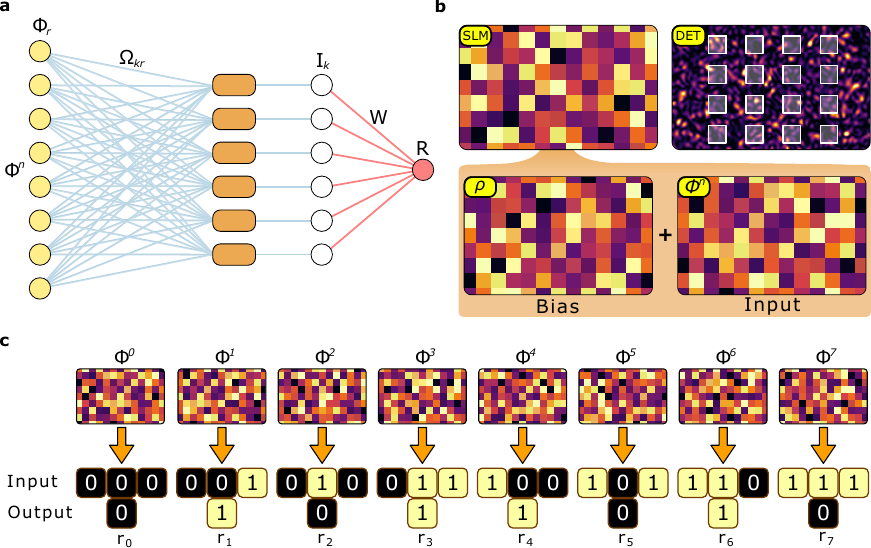}
  	\caption{Optical ELM implementation. {\bf a} ELM architecture. The pixel values ($\phi_r$) of each one of the input masks ($\phi^n$) are connected with random, fixed weights ($\Omega_{kr}$) to the hidden layer (with node values $I_k$). This hidden layer is connected to the output ($R$) via an output layer, $W$, which can be trained digitally simply by using linear regression. {\bf b} ELM implementation. In order to create random weights for the ELM, the SLM encodes the sum of two masks: a random, fixed bias mask($\rho$), which is used to create a rich speckle pattern that fills the whole sensor, and a unique input mask ($\phi^n$) for each input of the system. The detector measures the intensities at a given number of nodes (marked by the white squares), which can be chosen by the user and the complexity of the task to perform. {\bf c} Codification strategy for a simple CA (ECA rule 90). For each input of the table of truth (3-cell neighborhood), a unique phase mask is loaded onto the SLM. The resulting intensities measured by the camera are mapped to the desired outputs ($r_n$) via a digital output layer ($W$).}
  	\label{fig:fig2}
\end{figure}

A fundamental ELM consists of feedforward ANN with a fully-connected single layer or multiple layers of hidden nodes (see Fig. \ref{fig:fig2}a), and can be used for classification, clustering, and feature learning, among other tasks \cite{wangReviewExtremeLearning2022}. Contrary to conventional ANN's, the weights of the hidden layer can be randomly chosen and remain fixed during the whole training procedure. In order to optimize the system, only the weights between the hidden layer and the output layer are trained using a least-squares approach. This unique feature differentiates the training of ELM's from other ANN's, as there is no need to perform the (often demanding) training via the back-propagation algorithm. Experimentally, this is also beneficial, as the training can be done in-situ without the need of implementing the optical system in a simulation, which is computationally costly and tends to introduce errors due to optical aberrations, miss-alignments, and mismatches between digital models and real-world systems \cite{momeniBackpropagationfreeTrainingDeep2023}.

As explained before, the weights of our computational nodes are fixed and given by the geometry of the optical system. However, ELM's typically use random weights. Here, we show how it is possible to convert the fixed weights determined by the kernel of the FT into random weights without any change to the optical system, and thus implement an optical version of an ELM. For simplicity, we assume homogeneous illumination of the SLM, and phase-only modulation to codify the system inputs (i.e., the inputs of the system are encoded on a phase mask, $\phi$). Thus, without further modifications, the intensity at a given pixel of the detector for the $n$-th input phase mask ($\phi^{n}$) will be given by:

\begin{equation}
I_k^n = \left|\left| \sum_{r}{w_{k r}e^{i\phi_r^n}} \right|\right|^2,
\label{eq:discrete_node_constant_amplitude}
\end{equation}

where, again, the index $k$ ($r$) is used to specify pixel positions on the detector (SLM). It is straightforward to convert the weights into random values by codifying a fully random bias phase mask on the SLM ($\rho(x,y) \,$), which is added to the phase masks used to codify the inputs of the system. In this scenario, the intensity at a single detector position is given by:

\begin{equation}
I_k^n = \left|\left| \sum_{r}{w_{k r}e^{i(\phi_r^n + \rho_r)}} \right|\right|^2
	=  \left|\left| \sum_{r}{\Omega_{k r}e^{i\phi_r^n }} \right|\right|^2,
\label{eq:ELM_node}
\end{equation}

where we have introduced $\Omega_{k r} = w_{k r}e^{i \rho_r} = e^{-i[\frac{2\pi}{\lambda f}(xx' + yy') - \rho(x,y) ]} $, now a random weight. Optically, the whole system's output can be understood as the interference between the propagated field coming from the random phase mask (which will result in a random speckle pattern) and the phase mask codifying the input to the system (see Fig. \ref{fig:fig2}b). Thus, the intensity at each pixel of the detector is the result of a unique random projection of the input coded on the SLM. Moreover, the fact that the information is encoded on the complex phase of the field and, to a lesser extent, that the system works by measuring intensity, introduces the required non-linearities to implement complex operations. In theory, such a system would be able to implement as many random projections in parallel as the number of pixels of the detector, which can be on the order of several millions in modern digital cameras. In practice, however, the number of computation nodes is determined by the ratio between the sensor area and the speckle grain size, as measuring on pixels inside the same speckle grain will not yield different information for the same input mask (see Materials and Methods for an estimation in our system). Mathematically, the system operation can be modeled by the following input-output relationship. For a given input, $\phi^n$, the system returns a set of $M$ intensities:

\begin{equation}
\vec{I_n}=[I_{n0}, I_{n1}, ...\, , I_{nM}] = F\{\phi^n\},
\label{eq:input_output}
\end{equation}

where the operator $F$ represents the forward model of the optical system (codification on the complex phase of the field via the SLM, light propagation, and the modulus square operation carried out by the detector), and $M$ is the number of chosen detector regions (nodes). Ideally, one would want that, for a given input, the output of the system would directly match the output of the desired operation. In practice, though, the field at the detector is a speckle pattern, and the recorded intensities are random. However, while random, the system's response is fully deterministic, and different inputs will generate totally different (and linearly independent) output intensity vectors. For a set of $N$ inputs ($\vec{\phi} = [\phi^0, \phi^1, ...\,, \phi^N]$), the outputs of the system can be written in matrix form as:

\begin{equation}
	F\{\vec{\phi}\} = O =
	\begin{pmatrix}
	I_{00} & I_{01} & \hdots & I_{0M} \\
	I_{10} & I_{11} & \hdots & I_{1M} \\
	\vdots & \vdots&  \ddots & \vdots \\
	I_{N0} & I_{N1} & \hdots &I_{NM} \\
	\end{pmatrix}
	,
\label{eq:intensities_matrix_form}
\end{equation}

where the output vectors are arranged as the rows of matrix $O$. Then, for the output to match a desired operation, one can add a readout layer, $\vec{W}$, that converts the intensities into the desired result, $\vec{R}$:

\begin{equation}
	\vec{R} = O \cdot \vec{W} 
	\Rightarrow
	\begin{pmatrix}
	r_0 \\
	r_1 \\
	\vdots \\
	r_N 
	\end{pmatrix} 
	=
	\begin{pmatrix}
	I_{00} & I_{01} & \hdots & I_{0M} \\
	I_{10} & I_{11} & \hdots & I_{1M} \\
	\vdots & \vdots&  \ddots & \vdots \\
	I_{N0} & I_{N1} & \hdots &I_{NM} \\
	\end{pmatrix}
	\cdot 
	\begin{pmatrix}
	w_0 \\
	w_1 \\
	\vdots \\
	w_M 
	\end{pmatrix}
	.
\label{eq:readout_layer}
\end{equation}

Here, each element of $\vec{R}$ ($r_n$) can be, for example, one of the outputs of a table of truth for a given CA. In Fig. \ref{fig:fig2}c we show an example of this codification scheme for a simple CA (ECA rule 90). In order to obtain the elements of $\vec{W}$, one just simply needs to build $N$ distinct phase masks for each of the $N$ possible inputs of the desired operation ($2^3 = 8$ in the case of ECA, as the evolution of the automata is given by a cell and its two neighbors binary states), measure the resulting intensity vectors, and solve the equation system from Eq. \ref{eq:readout_layer}. For the system to have a solution, its rank has to be at least the same as the number of possible inputs of the operation to be implemented with the ELM ($N$). Thus, the number of measurement regions on the camera ($M$) has to be at least equal to $N$. While this is a sufficient condition in theory, in practice we generate speckle patterns on the detector plane. Given their nature, the intensity statistics of the speckle images follows a negative exponential. If $M$ is small, there is a non-zero chance that all the measurement regions will be on regions of the sensor with no light, which would reduce the rank of the equation system.  An easy fix to this is to measure in $M>N$ regions, with the size of each region being bigger than the speckle grain size, and then solve an over-determined equation system by using the Moore-Penrose pseudoinverse of $O$: $\vec{W} = O^{+}\vec{R}$. In principle, the only requirements that the masks need to fulfill is that they generate different intensities on the detector (i.e., that the rows of $O$ are linearly independent). We choose to build them simply by using random phase masks, but other approaches, like using elements of an orthogonal basis (e.g., Fourier, Haar, or Hadamard) should also work. Once this step has been carried out, mapping the system's output for a given input to the desired operation's output is done via $r_n = \vec{I_n}\cdot \vec{W} $. Following this method, it is possible to simulate different CA and observe a wide range complex emergent phenomena, as we show next.

\subsection*{Simulating emerging complex phenomena using CA}

\begin{figure}[h!]
	\centering
  	\includegraphics[width = 0.9\linewidth]{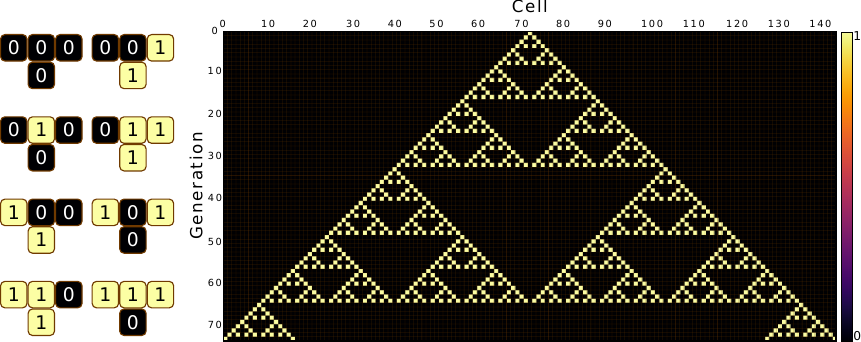}
  	\caption{ECA simulation using an optical ELM. ECA rule 90 table of truth (left) and its evolution for a starting seed consisting of a single alive cell at the center of the lattice. The pattern matches the Sierpinski Triangle, a fractal structure.}
  	\label{fig:fig3}
\end{figure}

In a first experiment, we implement the most basic kind of CA, labelled by Wolfram as elementary cellular automata \cite{wolframStatisticalMechanicsCellular1983}. ECA consist of a one-dimensional lattice of cells with binary states that evolve according to the following expression:

\begin{equation}
x_i(t+1) = f( x_{i-1}(t),\, x_i(t),\, x_{i+1}(t)),
\label{eq:ECA_rule}
\end{equation}

where $x_i(t)$ corresponds to the binary state (0 or 1, dead or alive) of the $i$-th cell at timestep $t$, and $f$ represents the update rule. It has to be noted that the update rules specifying cell interactions are applied by using only the cell states of a cell and its immediate neighbors, without taking into consideration the global state of the lattice, and thus there exist 256  ($2^{2^3}$) possible ECA rules, which encapsulate a wide range of behaviors. Experimentally, we implement any ECA in a time multiplexed fashion. Once a lattice size has been defined, we run along the initial grid states in groups of three cells, and calculate the state of the central cell for the next generation using the corresponding table of truth that encodes the given ECA rule (see Visualization 1 for a conceptual view of the evolution of the system). After the whole lattice has been covered, the calculated states form the next generation of the grid, a process that can be repeated sequentially for as long as desired. Fig. \ref{fig:fig3} shows both the table of truth for ECA rule 90 and its evolution with an initial lattice that consists of 145 cells, with a single alive cell, over 72 generations. It is possible to see that the diagram follows the Sierpinski Triangle, which can be constructed by recursively dividing an equilateral triangle into four equilateral smaller triangles, and removing the central one.

\begin{figure}[h!]
	\centering
  	\includegraphics[width = 0.8\linewidth]{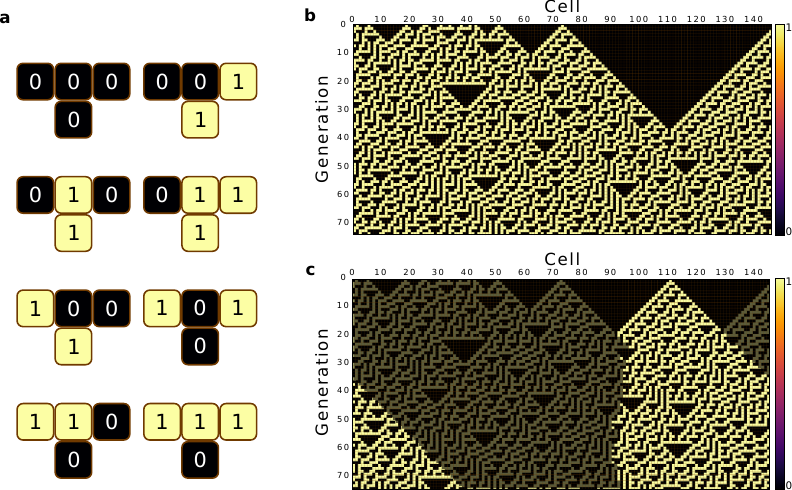}
  	\caption{Emergent chaotic behavior from a CA. {\bf a} Truth table for ECA rule 30 {\bf b} Evolution of the lattice showing chaotic dynamics over time. {\bf c} If a single cell is changed on the initial state, the pattern that is formed by the system presents differences that grow over time. Regions with the same state as in {\bf b} are shadowed, while cells with different states are highlighted. }
  	\label{fig:fig4}
\end{figure}

While rule 90 can be used to demonstrate fractal behavior, different rules illustrate other complex emerging properties of CA. In Fig. \ref{fig:fig4}, we show the evolution of ECA rule number 30, which belongs to a group of CA able to generate seemingly-random and chaotic patterns (class IV automata) \cite{wolframUniversalityComplexityCellular1984}. To highlight this chaotic behavior, we present the evolution of the system for two initial configurations which differ in the state of a single cell. For a given initial lattice arrangement, we let the system evolve over time, creating the diagram shown in Fig. \ref{fig:fig4}b. If a single cell of the initial seed is inverted, the system starts showing a clear differentiation that grows over time. This can be seen in Fig. \ref{fig:fig4}c, where the single cell perturbation introduced in the initial seed propagates laterally over the evolution of the system. Given our implementation with periodic boundary conditions, the perturbation keeps propagating at the left part of the diagram once it reaches the ending region of the lattice around generation 40. In the Supplement, we attach the evolution of the system for different ECA rules.

\begin{figure}[h]
	\centering
  	\includegraphics[width = 0.9\linewidth]{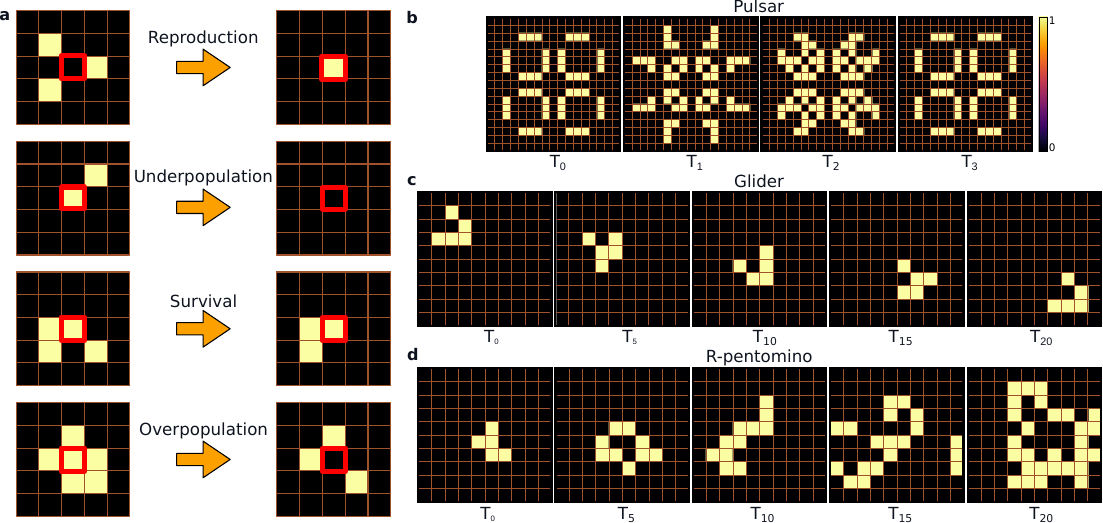}
  	\caption{Optical ELM implementation of Conway's Game of Life. {\bf a} The four update rules for CGoL. If a dead cell is surrounded by exactly three alive cells, it becomes alive at the next generation, as if by reproduction. Any alive cell with less than two alive neighbors dies (underpopulation). Any live cell with two or three alive neighbors survives. Any live cell with more than three live neighbors dies (overpopulation). We show these four rules for the central cell of a $5 \times 5$ grid, highlighted in red. {\bf b} Evolution of the Pulsar oscillator for four generations on a $17 \times 17$ lattice. Glider spaceship ({\bf c}) and R-pentomino Methuselah ({\bf d}) on a $10 \times 10$ lattice, showing system's evolution every 5 generations.}
  	\label{fig:fig5}
\end{figure}

The flexibility of our system can also be exploited to implement more complex CA without any experimental modifications. Instead to limiting ourselves to 1D CA, it is possible to implement higher dimensional automata, such as Conway's Game of Life (CGoL), by just increasing the table of truth size of the operations that the system performs. Devised by Conway in the 70's, CGoL is played on a 2D square grid where each lattice cell can have two different states (dead or alive, 1 or 0). Every cell interacts with its 8 neighbors following a simple set of rules (see Fig. \ref{fig:fig5}a), and the evolution of the system is calculated from generation to generation by applying the rules to all the cells of the lattice at the same time. Given an initial seed for the lattice, CGoL shows the emergent creation of patterns, groups of cells that evolve following different behaviors: still lifes, which remain unchanged along different generations; oscillators, which periodically return to an initial state after a given number of generations; and spaceships, which travel around the lattice. Similarly to the case of the ECA, after a 2D lattice has been defined, it is possible to run along its states in groups of $3\times 3$ cells calculating the states of the next generation by using a $2^9 = 512$ input table of truth (see Visualization 2 for a conceptual view of the system's evolution). In Fig. \ref{fig:fig5}b-d, we show the system evolution for a varied set of initial conditions a grid sizes, which re-create the pulsar (oscillator), glider (spaceship) and R-Pentomino methuselah. The full temporal evolution of these initial configurations can be seen on Visualizations 3-5.

\subsection*{Two-dimensional Turing machines}
To calculate the evolution of a CA, its rules are applied to all the cells of the lattice at the same time. If instead of that, one places a moving agent over the lattice and chooses update rules for both the lattice cells and the moving agent, the system becomes a two-dimensional version of a Turing machine (also called turmites, ants, or turning machines). Turmites can be understood as a small living termite (thus their name) living on an infinite 2D lattice of square cells. In its simplest form, the cells of the lattice have two possible states (0 or 1, dead or alive), and the turmite itself has two different properties: an orientation and a state. When the turmite travels around the lattice, it changes the states of the cells, which, as in the case of CA, can form intricate complex patterns \cite{langtonStudyingArtificialLife1986} and are capable of universal computation \cite{gajardoComplexityLangtonsAnt2002}. The system can be described with a state transition table, which contains all the possible inputs (cell and turmite state), and the corresponding outputs (new state for the cell, new state for the turmite, and the turmite rotation angle). An example table for Langton's ant, which reproduces both chaotic and order behavior, is shown in Table \ref{tab:ant}. At each time step, the ant has a given state and is looking at a given direction inside the cell. Then, it updates its state and direction, and moves one step forward, changing the state of the current cell (see Visualization 6 for a conceptual view of the system's evolution). These three values are simultaneously updated following the rules defined by the state transition table, and the procedure keeps repeating itself forever or until a stopping condition is met. 

\begin{table}[h]
\begin{center}
	\renewcommand{\arraystretch}{1.3}
	\resizebox{0.85\textwidth}{!}
		{
		\begin{tabular}{|c|c|c|c|c|c|c|c|}
			\multicolumn{2}{c}{\multirow{3}{*}{}} &
			\multicolumn{6}{c}{\textbf{Current cell state}} \\ \cline{3-8} 

			\multicolumn{2}{c|}{} &
			\multicolumn{3}{c|}{\textbf{0}} &
			\multicolumn{3}{c|}{\textbf{1}} \\ \cline{3-8} 
			
			\multicolumn{2}{c|}{ } &
			\makecell{\textbf{Set cell}\\\textbf{state to}} & \textbf{Turn} & \makecell{\textbf{Next}\\\textbf{ant state}} &
			\makecell{\textbf{Set cell}\\\textbf{state to}} & \textbf{Turn} & \makecell{\textbf{Next}\\\textbf{ant state}} \\ \hline
	
			\multirow{2}{*}{\makecell{\textbf{Current}\\\textbf{ant state}}}
			& \textbf{0}
			& 1 & R & 0		
			& 0 & L & 0 \\ \cline{2-8}	
			
			& \textbf{1}
			& 1 & R & 1		
			& 0 & L & 1 \\ \hline 		

		\end{tabular}
		}
	\caption{State transition table for Langton's ant. When arriving to a dead (0) cell, the ant turns right (R). If the cell is alive (1), the ant turns left (L). Then it updates its state for the next generation. In this simple case, the state of the ant does not really change the outcome of the system, but other turmite update rules depend on both the cell and turmite states.}
	\label{tab:ant}
\end{center}
\end{table}

\begin{figure}[h]
	\centering
  	\includegraphics[width = 0.6\linewidth]{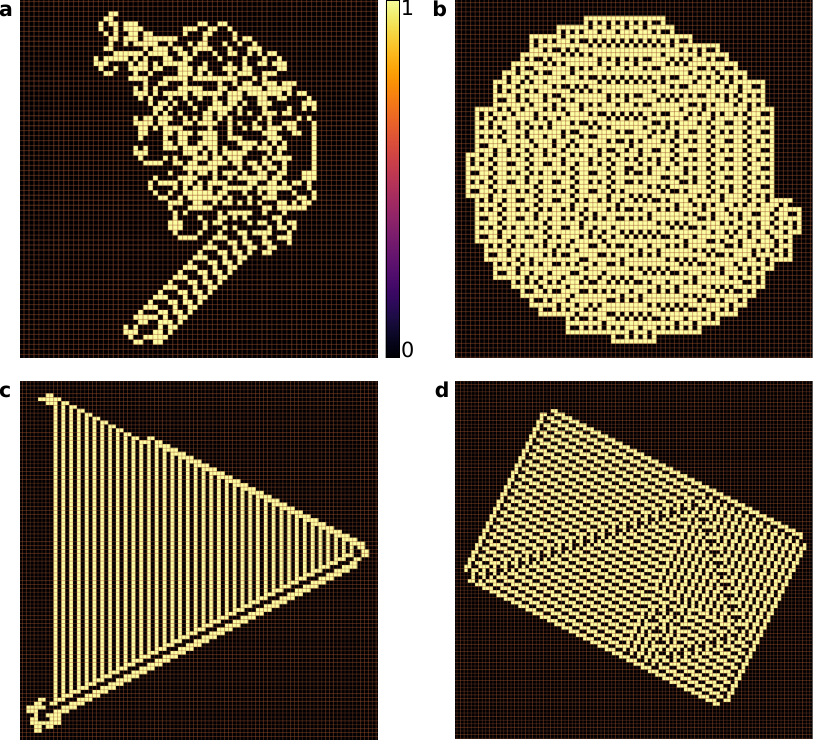}
  	\caption{Simulating turmites with an optical ELM. {\bf a} Langton's ant after $11000$ generations, showing chaotic (top) and order stages (bottom). {\bf b} Turmite generating spiral growth after 12536 generations. {\bf c} Turmite generating a triangular pattern after 10000 generations. {\bf d} Turmite showing a golden spiral pattern inscribed onto a rectangle filled with a periodic pattern after 10211 generations. All the turmites start with a totally empty lattice (0 state) and a single turmite.}
  	\label{fig:fig6}
\end{figure}

As in the case of ECA and CGoL, while the turmite movement and the cell state updates are determined by a small set of simple rules, and initially it could seem like the evolution is random, the whole system shows emergent complex phenomena. In Fig. \ref{fig:fig6}, we show four different turmite evolutions for a varied set of state transition tables. Langton's ant (Fig. \ref{fig:fig6}a) starts with a chaotic phase (top region of the lattice), but after $\sim$10000 steps, it enters a regular phase where it starts building a regular pattern commonly known as "highway" (bottom region of the lattice). Different state transition tables can be used to generate a wide variety of ordered patterns. For example, Figs. \ref{fig:fig6}b-d show the generation of patterns that match spiral growth, a triangle, and a rectangle that contains a golden spiral pattern. All the state transition tables for the generation of these patterns can be found in the Supplement, with images at different evolution stages. We also add multiple animations showing the turmites movements in Visualizations 7-10.

\section*{Discussion}

In this work, we have introduced a flexible strategy to build an affordable, experimentally simple, and efficient opto-electronic system for the implementation of CA. The system utilizes off-the-shelf optical components that are widely available in optics labs all over the world, its easy to build and control, compact, power efficient, very easy to align, and opens the possibility to face more complex computing tasks by harnessing the emerging complex behavior shown by CA \cite{liDeepLearningPhotonic2024}. It is fully programmable by design, and does not require any experimental modifications in order to simulate a wide range of artificial life systems. Moreover, its design does not require any complex optimization performed on a computer neither by using digital twins nor the back-propagation algorithm, as the whole training procedure can be carried directly in situ, and does not rely on the fabrication of nano/micro structured samples. While our current setup still relies in a digital computer to operate some of its parts (the training of the linear readout layer on the ELM implementation), this could be implemented directly in the electrical domain without the need to perform analog-to-digital conversion, thus creating a pure analog system that would operate even with a smaller energetic footprint and at faster speeds, and remain future research directions.

Given the high number of degrees of freedom available on the system (8-bit phase modulation of millions of pixels on the SLM and sensors with millions or even thousands of millions of pixels), our platform is a perfect candidate to implement massive parallelization of nonlinear random projections at the diffraction limit of light. Given the technology at its core, it is also well posed to take advantage of other parameters of light that can be used to encode and decode information, such as wavelength, polarization, and/or orbital angular momentum, all of which can be controlled by the use of different SLM technologies. Last, the use of neuromorphic-based detectors, such as event cameras \cite{wangEventCameraMeets2026} could open the gate to implement different kinds of ANN systems such as spiking neural networks \cite{feldmannAllopticalSpikingNeurosynaptic2019, xuSparseTransmissionDiffractive2026}, which are expected to decrease even further the power requirements and speed-up computing tasks for more complex CA. Along these lines, while the current results have been focused on the simulation of discrete CA and 2D Turing machines, the capabilities of free space optical systems to perform FT and convolution operations indicate that this approach could be an efficient tool for the simulation of long-range or continuous CA, whose behavior becomes closer to recurrent convolutional neural networks \cite{gilpinCellularAutomataConvolutional2019, chanLeniaExpandedUniverse2020}, but are much more computationally demanding than discrete CA.

\section*{Materials and Methods}

\subsection*{Experimental setup}

Light coming from a CW He-Ne laser ($\lambda = 632.8$ nm, 965867, JDS Uniphase Corporation) was collimated and expanded to fill the area of a LCoS-SLM (X15213-01, Hamamatsu Photonics K.K.). The SLM modulated the incoming beam, which was directed to the camera (acA3088-57um, Basler) passing through a single lens. The SLM-lens and lens-camera distances were the same as the focal length of the lens, which made the field distribution at the detector plane to correspond to the FT of the field at the SLM plane. Custom Python codes were used to control the SLM and the camera, the generation of the phase masks, and the calculation and application of the weights of the output layer of the ELM. They can be found at Ref. \cite{morenoCodesRawData2026}.

\subsection*{Measurement acquisition }
To avoid the crosstalk effect between the pixels of the SLM, the random masks had a binning of $4\times 4$ pixels. The recorded images had a size of $2064 \times 3088$ $px$, with an integration time of 4000 $\mu$s. In all the measurements, the total optical power at the camera plane was 1.930 $\mu$W. Given the refresh rate of the SLM (60 Hz) and the low integration times required, each measurement was done with an average of three camera images. In our experiments for ECA ($N = 8$) and turmite ($N=4$) simulations, a grid of $10 \times 10$ detection regions was chosen, which proved to be high enough so the equation system from Eq. \ref{eq:readout_layer} always had a solution. In the case of CGoL ($N = 512$), a grid of $30 \times 30$ was used. In all cases, each measurement region was chosen so its size ($40 \times 40$ px) was bigger than the speckle grain size ($20 \times 20$ px) to avoid close range correlations and small vibrations effects. With these experimental conditions, the maximum number of output nodes for the ELM was on the order of 4000.

\section*{Data availability}

The data and codes that support the plots and results of the manuscript can be found at Ref. \cite{morenoCodesRawData2026}.

\section*{Acknowledgments}
We acknowledge Project No. CNS2023-145510 funded by MCIN/AEI/10.13039/501100011033, “European Union NextGenerationEU/PRTR”, Project No. CIPROM/2023/44 funded by Generalitat Valenciana, and Project No. PID2024-158832NB-C22. This work was supported by DYNAMO project (101046489), funded by the European Union. Views and opinions expressed are however those of the authors only and do not necessarily reflect those of the European Union or European Innovation Council. Neither the European Union nor the granting authority can be held responsible for them.

\section*{Contributions}
D.T. and F.S. conceived the project. E.M and F.S. developed the optical setup, with inputs from D.T.  E.M. performed the experiments and analyzed the results. E.M. and F.S. wrote the manuscript with inputs from all the authors. D.T. and F.S. supervised the project. D.T. secured funding for the project. 

\bibliographystyle{unsrt}
\bibliography{bibliography}

\end{document}